\documentclass[10pt,english]{revtex4}
\usepackage{lmodern}

\usepackage[T1]{fontenc}
\usepackage[latin9]{inputenc}
\usepackage{amsmath}
\usepackage{graphicx}
\usepackage{amssymb}
\usepackage{esint}

\makeatletter
\@ifundefined{textcolor}{}
{%
 \definecolor{BLACK}{gray}{0}
 \definecolor{WHITE}{gray}{1}
 \definecolor{RED}{rgb}{1,0,0}
 \definecolor{GREEN}{rgb}{0,1,0}
 \definecolor{BLUE}{rgb}{0,0,1}
 \definecolor{CYAN}{cmyk}{1,0,0,0}
 \definecolor{MAGENTA}{cmyk}{0,1,0,0}
 \definecolor{YELLOW}{cmyk}{0,0,1,0}
 }

\usepackage{latexsym}\usepackage{bm}

\makeatother

\makeatother

\usepackage{babel}

\makeatother

\usepackage{babel}

\begin{document}

\title{c-functions in the Born-Infeld extended new massive gravity}

\author{\.{I}brahim Güllü }

\email{e075555@metu.edu.tr}

\affiliation{Department of Physics,\\
 Middle East Technical University, 06531, Ankara, Turkey}

\author{Tahsin Ça\u{g}r\i{} \c{S}i\c{s}man}

\email{sisman@metu.edu.tr}

\affiliation{Department of Physics,\\
 Middle East Technical University, 06531, Ankara, Turkey}

\author{Bayram Tekin}

\email{btekin@metu.edu.tr}

\affiliation{Department of Physics,\\
 Middle East Technical University, 06531, Ankara, Turkey}

\date{\today}
\begin{abstract}
We derive and study the equations of motion of the Born-Infeld extension
of new massive gravity for globally and asymptotically (anti-)de Sitter
spaces, and show that the assumptions of the null-energy condition
and holography (that bounds the $c$-function) lead to two simple
$c$-functions, one of which is equivalent to the $c$-function of
Einstein's gravity. We also show that, at the fixed point, the $c$-function
gives the central charge of the Virasoro algebra and the coefficient
of the Weyl anomaly up to a constant.
\end{abstract}
\maketitle

\section{Introduction}

New massive gravity (NMG) that provides a nonlinear extension to the
Pauli-Fierz massive gravity was introduced in \cite{BHT} with the
action \begin{equation}
I_{\text{NMG}}=\frac{1}{\kappa^{2}}\int d^{3}x\,\sqrt{-\det g}\left[\sigma R+\frac{1}{m^{2}}\left(R_{\mu\nu}^{2}-\frac{3}{8}R^{2}\right)-2\lambda m^{2}\right].\label{eq:NMG_action}\end{equation}
 At the linearized level, this theory is unitary for both flat and
(a)dS spaces \cite{BHT,Nakasone,Deser,Gullu1,Gullu2} for $\sigma=-1$.
The question arises as to what possible additional higher-derivative
terms beyond the quadratic ones can be added to this action keeping
the unitarity intact. In flat spaces, at the linearized level, such
terms will not change the unitarity property. But, for other backgrounds
one has to find a guiding principle, since higher-derivative terms
could enter in many different forms and combinations. In \cite{Sinha},
AdS/CFT correspondence along with the existence of a holographic $c$-theorem
was used to extend NMG to $O\left(R^{3}\right)$ and $O\left(R^{4}\right)$.
At each order, a new coupling constant is introduced with this procedure.
In \cite{Gullu3}, another proposal of extending NMG to all orders
was made which is based on a Born-Infeld type action, and reads as\begin{equation}
I_{\text{BI-NMG}}=-\frac{4m^{2}}{\kappa^{2}}\int d^{3}x\,\left[\sqrt{-\det\left(g+\frac{\sigma}{m^{2}}G\right)}-\left(1-\frac{\lambda}{2}\right)\sqrt{-\det g}\right],\label{eq:BI-NMG_action}\end{equation}
 where $G_{\mu\nu}\equiv R_{\mu\nu}-\frac{1}{2}g_{\mu\nu}R$ and $\sigma=\pm1$.
What is quite interesting about (\ref{eq:BI-NMG_action}) is that
at the first order it gives the cosmological Einstein-Hilbert action,
at the second order it gives the NMG action, and at the third and
fourth orders it gives the action introduced in \cite{Sinha} upon
fixing the arbitrary coupling constants in the latter. See Appendix
A for $O\left(R^{4}\right)$ matching of (\ref{eq:BI-NMG_action})
with that of \cite{Sinha}, $O\left(R^{3}\right)$ matching was already
shown in \cite{Gullu3}. We expect that this matching will work at
any order.

BI type actions, with their singularity-free solutions and due to
their nice properties such as ghost freedom, causal propagation, have
appeared in physics ranging from Maxwell theory to gravity and string
theory. The fact that with such a simple action (\ref{eq:BI-NMG_action})
one can describe nonlinearly massive gravity, albeit in three dimensions,
is remarkable. As we shall see below, what is also interesting with
the BI-NMG is that, unlike the NMG, for a given viable $\lambda$
there is a unique vacuum solution with an effective cosmological constant
$\Lambda$. In the case of NMG and other finite order deformations
of it with higher curvatures, one has to deal with degenerate vacua
{[}two for NMG with cosmological constants $\Lambda_{\pm}=-2m^{2}\left(\sigma\pm\sqrt{1+\lambda}\right)${]}.
In quantum field theory, degenerate vacua would normally cause no
problem, but for gravity it is not clear at all which vacuum should
be chosen: as far as energy is concerned, one cannot compare spaces
which asymptote to different constant curvature manifolds. {[}Note
that in NMG, at $\lambda=-1$, degeneracy is lifted and one has a
unique vacuum. See the solutions of this theory \cite{BHT,Troncoso}.
But, clearly $\lambda=-1$ is a very special point.{]} 

In this work, we find the field equations of BI-NMG, and study the
constant curvature and asymptotically constant curvature spaces. More
specifically, with the assumption of the null-energy condition %
\footnote{For $\sigma=+1$, one also needs the assumption that the space is
asymptotically AdS.%
}, we show that the BI-NMG action has two simple $c$-functions. One
of these $c$-functions is just like the $c$-function of Einstein's
gravity and at infinity coincides with the central charge of the Virasoro
algebra and the Weyl anomaly coefficient up to a constant. 

The layout of the paper is as follows. In Section II, we give the
equations of motion for the exact BI-NMG theory and find constant
curvature solutions. In Section III, we investigate the $c$-function
of the BI-NMG theory, and compute the central charge and the Weyl
anomaly coefficient. Some of the computations are expounded upon in
the appendices.

\section{Equations of Motion and Constant Curvature Solutions}

To find the equations of motion it is somewhat better (if less compact)
to expand the determinant with the help of \begin{equation}
\det A=\frac{1}{6}\left[\left(\text{Tr}A\right)^{3}-3\text{Tr}A\text{Tr}\left(A^{2}\right)+2\text{Tr}\left(A^{3}\right)\right],\label{eq:detA_in_traces}\end{equation}
 which is valid for a generic $3\times3$ matrix. Then, one computes
\begin{align}
\text{Tr}\left(1+\frac{\sigma}{m^{2}}g^{-1}G\right) & =3-\frac{\sigma}{2m^{2}}R,\qquad\text{Tr}\left(g^{-1}G\right)^{2}=R_{\mu\nu}^{2}-\frac{1}{4}R^{2},\nonumber \\
\text{Tr}\left(g^{-1}G\right)^{3} & =R^{\mu\nu}R_{\nu}^{\phantom{\nu}\alpha}R_{\alpha\mu}-\frac{3}{2}RR_{\mu\nu}^{2}+\frac{3}{8}R^{3},\label{eq:Traces}\end{align}
 which reduces the action to \cite{Gullu4} \begin{equation}
I_{\text{BI-NMG}}=-\frac{4m^{2}}{\kappa^{2}}\int d^{3}x\,\sqrt{-\det g}F\left(R,K,S\right),\label{eq:BI-NMG_trace}\end{equation}
 where\begin{equation}
F\left(R,K,S\right)\equiv\sqrt{1-\frac{\sigma}{2m^{2}}\left(R+\frac{\sigma}{m^{2}}K-\frac{1}{12m^{4}}S\right)}-\left(1-\frac{\lambda}{2}\right),\label{eq:F_defn}\end{equation}
 \begin{equation}
K\equiv R_{\mu\nu}^{2}-\frac{1}{2}R^{2},\qquad S\equiv8R^{\mu\nu}R_{\mu\alpha}R_{\phantom{\alpha}\nu}^{\alpha}-6RR_{\mu\nu}^{2}+R^{3}.\label{eq:K_S_defns}\end{equation}
 After a quite lengthy calculation which we partly give in Appendix
B, one arrives at\begin{align}
-\frac{\kappa^{2}}{8m^{2}}T_{\mu\nu} & =-\frac{1}{2}Fg_{\mu\nu}+\left(g_{\mu\nu}\Box-\nabla_{\mu}\nabla_{\nu}\right)F_{R}+F_{R}R_{\mu\nu}\nonumber \\
 & \phantom{=}-\frac{\sigma}{m^{2}}\left\{ 2\nabla_{\alpha}\nabla_{\mu}\left(F_{R}R_{\phantom{\alpha}\nu}^{\alpha}\right)-g_{\mu\nu}\nabla_{\beta}\nabla_{\alpha}\left(F_{R}R^{\alpha\beta}\right)-\Box\left(F_{R}R_{\mu\nu}\right)-2F_{R}R_{\nu}^{\phantom{\nu}\alpha}R_{\mu\alpha}\right.\nonumber \\
 & \phantom{=-\frac{\sigma}{m^{2}}}\left.+g_{\mu\nu}\Box\left(F_{R}R\right)-\nabla_{\mu}\nabla_{\nu}\left(F_{R}R\right)+F_{R}RR_{\mu\nu}\right\} \nonumber \\
 & \phantom{=}-\frac{1}{2m^{4}}\Biggl\{4F_{R}R_{\phantom{\rho}\mu}^{\rho}R_{\rho\alpha}R_{\phantom{\alpha}\nu}^{\alpha}+2g_{\mu\nu}\nabla_{\alpha}\nabla_{\beta}\left(F_{R}R^{\beta\rho}R_{\phantom{\alpha}\rho}^{\alpha}\right)+2\Box\left(F_{R}R_{\nu}^{\phantom{\nu}\rho}R_{\mu\rho}\right)\label{eq:Eqn_of_motn}\\
 & \phantom{=-\frac{1}{2m^{4}}}-4\nabla_{\alpha}\nabla_{\mu}\left(F_{R}R_{\nu}^{\phantom{\nu}\rho}R_{\phantom{\alpha}\rho}^{\alpha}\right)+2\nabla_{\alpha}\nabla_{\mu}\left(F_{R}RR_{\phantom{\alpha}\nu}^{\alpha}\right)-g_{\mu\nu}\nabla_{\alpha}\nabla_{\beta}\left(F_{R}RR^{\alpha\beta}\right)\nonumber \\
 & \phantom{=-\frac{1}{2m^{4}}}-\Box\left(F_{R}RR_{\mu\nu}\right)-2F_{R}RR_{\nu}^{\phantom{\nu}\rho}R_{\mu\rho}-g_{\mu\nu}\Box\left(F_{R}R_{\alpha\beta}^{2}\right)+\nabla_{\nu}\nabla_{\mu}\left(F_{R}R_{\alpha\beta}^{2}\right)\nonumber \\
 & \phantom{=-\frac{1}{2m^{4}}}-F_{R}R_{\alpha\beta}^{2}R_{\mu\nu}+\frac{1}{2}g_{\mu\nu}\Box\left(F_{R}R^{2}\right)-\frac{1}{2}\nabla_{\mu}\nabla_{\nu}\left(F_{R}R^{2}\right)+\frac{1}{2}F_{R}R^{2}R_{\mu\nu}\Biggr\},\nonumber \end{align}
 where\begin{equation}
F_{R}=\frac{\partial F}{\partial R}=-\frac{\sigma}{4m^{2}\left[F+\left(1-\frac{\lambda}{2}\right)\right]}.\label{eq:F_R_defn}\end{equation}

Clearly, flat spacetime is a solution for these equations when $\lambda=0$,
since the Riemann tensor is zero which yields $F=0$, and $F_{R}=-\frac{\sigma}{4m^{2}}$.
In order to find the constant curvature solutions, let \begin{equation}
R_{\mu\rho\nu\sigma}=\Lambda\left(g_{\mu\nu}g_{\rho\sigma}-g_{\mu\sigma}g_{\nu\rho}\right),\qquad R_{\rho\sigma}=2\Lambda g_{\rho\sigma},\qquad R=6\Lambda.\label{eq:curv_tensors}\end{equation}
 With these, $F$ becomes\begin{align}
F\left(R,K,S\right) & =\sqrt{\left(1-\frac{\sigma\Lambda}{m^{2}}\right)^{3}}-\left(1-\frac{\lambda}{2}\right),\label{eq:F_in_const_curv_back}\end{align}
 where the cosmological constant is restricted as $\frac{\sigma}{m^{2}}\Lambda\le1$.
Putting these in the equation of motion, one obtains\begin{equation}
0=\left(-\frac{1}{2}F-\frac{\sigma\Lambda}{2m^{2}\left[F+\left(1-\frac{\lambda}{2}\right)\right]}+\frac{\Lambda^{2}}{m^{4}\left[F+\left(1-\frac{\lambda}{2}\right)\right]}-\frac{\sigma\Lambda^{3}}{2m^{6}\left[F+\left(1-\frac{\lambda}{2}\right)\right]}\right)g_{\mu\nu}.\label{eq:const_back_eom}\end{equation}
 which puts another constraint $\sigma\frac{\Lambda}{m^{2}}\ne1$.
Let us define $x\equiv-\sigma\frac{\Lambda}{m^{2}}$, then the above
equation can be put in the following form\begin{equation}
\left(1+x\right)^{2}=\left(1-\frac{\lambda}{2}\right)\sqrt{\left(1+x\right)^{3}}.\label{eq:x_eqn}\end{equation}
 which restricts $\lambda$ as $\lambda<2.$ Here, note that $\left(1-\frac{\lambda}{2}\right)$
cannot be equal to zero, since it implies $\sigma\frac{\Lambda}{m^{2}}=1$
which then leads to vanishing $F$. It is important to note that BI-NMG
theory does not restrict $\lambda$ for generic, that is nonmaximally
symmetric, spaces. Solving (\ref{eq:x_eqn}) gives \begin{equation}
\Lambda=\sigma m^{2}\lambda\left(1-\frac{\lambda}{4}\right),\qquad\lambda<2.\label{eq:Lambda_eqn}\end{equation}
 Unlike the NMG case, as we mentioned in the introduction, for a viable
$\lambda$, there is a \emph{unique} vacuum solution. For $\sigma=+1$,
$\Lambda$ and $\lambda$ have the same signs; for $\sigma=-1$, they
have the opposite signs. Minimum of $\Lambda$ occurs at $\lambda=2$,
but this point is not allowed. {[}Note that the discussion above works
exactly for static BTZ black holes: namely, they are solutions of
this theory.{]}

\section{$c$-functions of the BI-NMG theory}

\subsection{Central charges}

In order to discuss the degrees of freedom for the two-dimensional
CFT theory that is living on the boundary of $\text{AdS}_{3}$, let
us determine the central charges. In pure Einstein gravity in AdS,
the global $\text{SO}\left(2,2\right)$ symmetry is enlarged to two
copies of an infinite dimensional symmetry (Virasoro algebra) which
has the following central charge \cite{Brown} \begin{equation}
c=\frac{3\ell}{2G_{3}},\label{eq:central_charge_in_AdS-EH}\end{equation}
 where $\ell$ is the AdS length defined as $\Lambda\equiv-\frac{1}{\ell^{2}}$,
and $G_{3}$ is the three-dimensional Newton's constant which is related
to $\kappa$ as $G_{3}=\frac{\kappa^{2}}{16\pi}$. In understanding
(\ref{eq:central_charge_in_AdS-EH}), one has to be careful about
the role played by the symmetry and the role played by the dynamics,
that is the field equations. To get (\ref{eq:central_charge_in_AdS-EH}),
certain falloff conditions on how the space asymptotes $\text{AdS}_{3}$
are specified. Therefore, as the theory and the asymptotic falloff
conditions change, $c$ changes. For higher-derivative gravity theories,
central charge is given as \cite{Wald,Kraus,Saida,Imbimbo} \begin{equation}
c=\frac{\ell}{2G_{3}}g_{\mu\nu}\frac{\partial\mathcal{L}_{3}}{\partial R_{\mu\nu}},\label{eq:cent_charge_formula}\end{equation}
 where $\mathcal{L}_{3}$ in the BI-NMG gravity is $\mathcal{L}_{3}=-4m^{2}F\left(R,K,S\right).$
Then,\[
\frac{\partial\mathcal{L}_{3}}{\partial R_{\mu\nu}}=\frac{\sigma}{\left[F+\left(1-\frac{\lambda}{2}\right)\right]}\left\{ \left[1-\frac{\sigma}{m^{2}}R+\frac{1}{2m^{4}}\left(R_{\alpha\beta}^{2}-\frac{1}{2}R^{2}\right)\right]g^{\mu\nu}+\frac{2}{m^{2}}\left(\sigma+\frac{1}{2m^{2}}R\right)R^{\mu\nu}-\frac{2}{m^{4}}R_{\mu\alpha}R_{\phantom{\alpha}\nu}^{\alpha}\right\} ,\]
 and using (\ref{eq:curv_tensors}) with $\Lambda\equiv-\frac{1}{\ell^{2}}$,
the central charge becomes\begin{equation}
c=\frac{3\sigma\ell}{2G_{3}}\sqrt{1+\frac{\sigma}{\ell^{2}m^{2}}}=\frac{3\sigma}{4G_{3}\ell}\left(2-\lambda\right).\label{eq:BI-NMG_cent_chrg}\end{equation}
 where $\lambda<2$. For $c$ to be positive, $\sigma$ should be
positive: This seems to be in apparent conflict with the unitarity
of NMG which required $\sigma=-1$. But, one should still study the
bulk unitarity of $\sigma=+1$ theory in BI-NMG, which we have not
done yet.

\subsection{$c$-functions}

Let us recall that Zamolodchikov \cite{Zamolodchikov} proved, under
certain symmetry assumptions, that any nonconformal two-dimensional
theory has a so called $c$-function which monotonically decreases
under the renormalization group flow towards lower energies and matches
that of the central charge of the corresponding conformal field theory
that appears at the \emph{fixed} points of the flow. Beyond two dimensions,
it is hard to prove the existence of a $c$-theorem in general. But,
one can explicitly construct $c$-functions in certain theories \cite{Freedman}
that satisfy the null energy condition, $T_{\mu\nu}\zeta^{\mu}\zeta^{\nu}\ge0$,
where $\zeta$ is an arbitrary null vector. Following \cite{Freedman},
consider the domain wall ansatz:\begin{equation}
ds^{2}=e^{2A\left(r\right)}\left(-dt^{2}+dx^{2}\right)+dr^{2},\label{eq:domain_wall}\end{equation}
 and plug it to the field equations of pure Einstein gravity to get\begin{equation}
\frac{2A^{\prime\prime}}{\kappa^{2}}=T_{t}^{t}-T_{r}^{r}\le0.\label{eq:Einstein_dom_wall_inequ}\end{equation}
 The $c$-function can simply be defined as \begin{equation}
c\left(r\right)=\frac{24\pi}{\kappa^{2}A^{\prime}\left(r\right)}=\frac{3}{2G_{3}A^{\prime}\left(r\right)}.\label{eq:c-func_of_Einstein}\end{equation}
 With this definition, $c\left(r\right)$ satisfies a monotonic increase
towards higher energy, for increasing $r$. Its prefactor is arbitrary,
but one can fix it in such a way that one gets the central charge
of the Virasoro algebra that appears in the boundary of $\text{AdS}_{3}$.

Plugging (\ref{eq:domain_wall}) to (\ref{eq:Eqn_of_motn}), one arrives\begin{equation}
\left(\frac{2m}{\kappa^{2}}\right)\frac{\left[A^{\prime\prime}+\left(A^{\prime}\right)^{2}+\sigma m^{2}\right]A^{\prime\prime}}{\sqrt{\left[m^{2}+\sigma\left(A^{\prime}\right)^{2}\right]\left[A^{\prime\prime}+\left(A^{\prime}\right)^{2}+\sigma m^{2}\right]^{2}}}=T_{t}^{t}-T_{r}^{r}\le0,\label{eq:null_en_inequality}\end{equation}
 where $m$ is taken to be positive. In order to understand the constraints,
let us look at\begin{equation}
F_{R}=-\frac{\sigma m}{4\sqrt{\left[m^{2}+\sigma\left(A^{\prime}\right)^{2}\right]\left[A^{\prime\prime}+\left(A^{\prime}\right)^{2}+\sigma m^{2}\right]^{2}}},\label{eq:FR_for_dom_wall_anstz}\end{equation}
 which appears in the equation of motion. Finiteness and reality of
$F_{R}$ set two constraints\begin{equation}
\left[A^{\prime\prime}+\left(A^{\prime}\right)^{2}+\sigma m^{2}\right]\ne0,\qquad m^{2}+\sigma\left(A^{\prime}\right)^{2}>0.\label{eq:constaints_in_dom_wall}\end{equation}
 First of all, observe that the second constraint is automatically
satisfied for the $\sigma=+1$ case, but implies a bound for $A^{\prime}$
for the $\sigma=-1$ case. Let us discuss what the first constraint
implies: $\left[A^{\prime\prime}+\left(A^{\prime}\right)^{2}+\sigma m^{2}\right]$
cannot change sign for $r\in\left(-\infty,\infty\right)$. The sign
of $A^{\prime\prime}$ and the sign of $\left[A^{\prime\prime}+\left(A^{\prime}\right)^{2}+\sigma m^{2}\right]$
are correlated because of (\ref{eq:null_en_inequality}): namely,
for both values of $\sigma$, they should have the opposite signs.
Therefore, the sign of $A^{\prime\prime}$, also, cannot change. Hence,
$A^{\prime}$ is either monotonically increasing or decreasing. Let
us discuss the $\sigma=+1$ and $\sigma=-1$ cases separately:
\begin{itemize}
\item \textbf{\underbar{$\sigma=+1$ case}}\textbf{:} If one chooses $A^{\prime\prime}\ge0$,
then $\left[A^{\prime\prime}+\left(A^{\prime}\right)^{2}+m^{2}\right]$
is positive which conflicts with (\ref{eq:null_en_inequality}). So,
this choice is forbidden. The other option: $A^{\prime\prime}\le0$
leads to $A^{\prime\prime}+\left(A^{\prime}\right)^{2}+m^{2}>0$ which
only puts an upper bound on $\left|A^{\prime\prime}\right|$. Therefore,
$A^{\prime}$ is a monotonically decreasing function. To see the implications
of the second constraint, we should emphasize that the AdS/CFT correspondence
requires that the spacetime is asymptotically AdS such that as $r\rightarrow\infty$
(the UV region), $A\left(r\right)$ takes the form $A\left(r\right)=r\sqrt{\left|\Lambda\right|}$.
This gives a lower bound of $\sqrt{\left|\Lambda\right|}$ at $r\rightarrow\infty$
for $A^{\prime}$. Therefore, $A^{\prime}$ is positive at every $r$
. Then, if one assumes that the spacetime at $r\rightarrow-\infty$
(the IR region) is also an AdS spacetime, $A^{\prime}$ is bounded
from above. To sum up, for $\sigma=+1$, $A^{\prime\prime}\le0$ is
proven and $A^{\prime}$ is a monotonically decreasing function (a
possible form is depicted in the first plot of Fig. 1). It is important
to understand the role played by the AdS/CFT correspondence here:
without it we would still have $A^{\prime\prime}\le0$, but $A^{\prime}$
need not be bounded.\\
\begin{figure}
\includegraphics[scale=0.4]{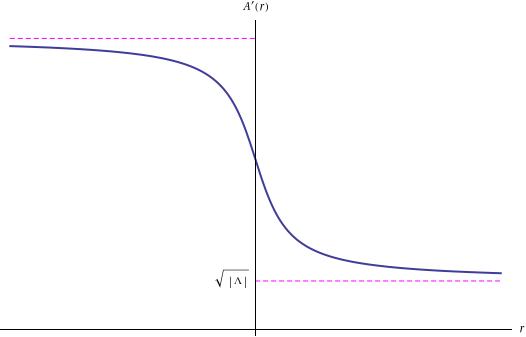}$\qquad\qquad$\includegraphics[scale=0.4]{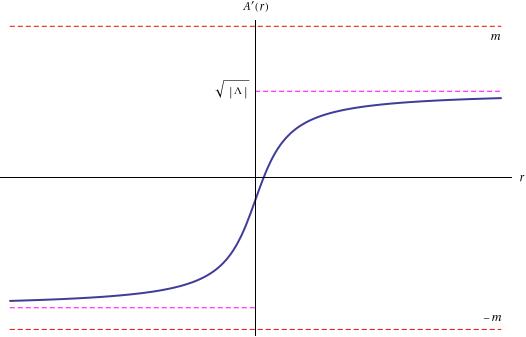}

\caption{The first graph gives a possible behavior for $A^{\prime}\left(r\right)$
in the $\sigma=+1$ case, while the second one shows a possible behavior
for $A^{\prime}\left(r\right)$ in the $\sigma=-1$ case.}

\end{figure}

\item \textbf{\underbar{$\sigma=-1$ case}}\textbf{:} The second constraint
can be written as $m>\left|A^{\prime}\right|>-m$. Monotonic and bounded
behavior of $A^{\prime}$ also requires that $A^{\prime\prime}=0$
as $r\rightarrow\pm\infty$. If one chooses $A^{\prime\prime}\le0$,
then $\left[A^{\prime\prime}+\left(A^{\prime}\right)^{2}-m^{2}\right]$
should be positive; however, this is not possible with $m^{2}-\left(A^{\prime}\right)^{2}>0$.
So, $A^{\prime\prime}\le0$ is ruled out. Consider the other case:
$A^{\prime\prime}\ge0$, which implies $m^{2}-\left(A^{\prime}\right)^{2}>A^{\prime\prime}$.
Therefore, $A^{\prime}$ is a monotonically increasing function. Here,
the fact that $A^{\prime}\left(+\infty\right)$ and $A^{\prime}\left(-\infty\right)$
are allowed to take positive and negative values, respectively, leads
to the interesting case of having two AdS boundaries (this means $A\left(r\right)\rightarrow\infty$).
Overall, for $\sigma=-1$, we proved that $A^{\prime\prime}\ge0$
and so $A^{\prime}$ is a monotonically increasing function (a possible
form is shown in the second plot of Fig. 1).
\end{itemize}
The above analysis shows that the assumption of the null-energy condition
leads to a simple $c$-function in the BI-NMG theory. As shown above,
for $\sigma=+1$, $A^{\prime\prime}\le0$ and for $\sigma=-1$, $A^{\prime\prime}\ge0$.
Then, the $c$-function for the BI-NMG theory can be defined as in
the Einstein's gravity: \begin{equation}
c\left(r\right)=\frac{3\sigma}{2G_{3}A^{\prime}\left(r\right)}\quad\Rightarrow\quad\frac{dc}{dr}\ge0,\label{eq:c-function}\end{equation}
 which at $r\rightarrow\infty$ gives the central charge of the Virasoro
algebra (\ref{eq:BI-NMG_cent_chrg}) (and the Weyl anomaly to be discussed
below) up to a factor. However, this is not the only possibility.
Another $c$-function can be determined by considering the inequality\begin{equation}
\frac{\sigma A^{\prime\prime}}{\sqrt{m^{2}+\sigma\left(A^{\prime}\right)^{2}}}\le0,\label{eq:inequlality_second}\end{equation}
 which is also implied by (\ref{eq:null_en_inequality}). Then, the
$c$-function is found by simply integrating the above form as\begin{equation}
c\left(r\right)=-\sigma\arctan\left(\frac{A^{\prime}\left(r\right)}{\sqrt{m^{2}+\sigma\left(A^{\prime}\right)^{2}}}\right),\label{eq:sec_c-func}\end{equation}
 which is also a monotonically increasing function. Note that a minus
sign is introduced to get the increasing behavior. The meaning of
this $c$-function (\ref{eq:sec_c-func}) in its exact form is not
immediately clear, but an expansion of (\ref{eq:inequlality_second})
around large $m^{2}$ gives at the desired order the $c$-functions
introduced by \cite{Sinha} (Note that at each order $\left(A^{\prime}\right)^{n}$
the $c$-functions of \cite{Sinha} have arbitrary parameters, while
these are fixed in our case.).

\subsection{Weyl anomaly}

The computation of the coefficient of Weyl anomaly can be done as
follows \cite{Skenderis,Emparan}. The Euclidean metric\begin{equation}
ds^{2}=\frac{dr^{2}}{1-\Lambda r^{2}}+r^{2}\left(d\theta^{2}+\sin^{2}\theta d\phi^{2}\right),\label{eq:Euclidean_metric}\end{equation}
 solves the equations of motion with $\Lambda=\sigma m^{2}\lambda\left(1-\frac{\lambda}{4}\right),\,\lambda<2.$
Here, note that AdS space is obtained for $\sigma=+1$ with the bound
$\lambda<0$, and for $\sigma=-1$ with the bound $2>\lambda>0$.
With this metric our action reads\begin{equation}
I_{\text{BI}}=\left(-\frac{4m^{2}}{\kappa^{2}}\right)\left[\left(1-\frac{\sigma\Lambda}{m^{2}}\right)^{3/2}-\left(1-\frac{\lambda}{2}\right)\right]\int d^{3}x\,\sqrt{\det g}.\label{eq:Euclid_act}\end{equation}
 Defining $\Lambda\equiv-\frac{1}{\ell^{2}}$, and putting a cutoff
($L$), one arrives at \begin{equation}
I_{\text{BI}}=\left(\frac{16\pi m^{2}}{\kappa^{2}\ell}\right)\lambda\left(1-\frac{\lambda}{4}\right)\left(1-\frac{\lambda}{2}\right)\frac{L^{2}}{2}\left[\sqrt{\left(\frac{\ell}{L}\right)^{2}+1}-\left(\frac{\ell}{L}\right)^{2}\text{arcsinh}\left(\frac{L}{\ell}\right)\right].\label{eq:Euc_act_after_int}\end{equation}
 which is valid for both $\sigma=+1$ and $\sigma=-1$ {[}Note that
$\text{sign}\left(\lambda\right)=-\sigma${]}. As $L\rightarrow\infty$,
after dropping a quadratic divergence, the action becomes\begin{align}
I_{\text{BI}} & \approx-\left(\frac{8\pi m^{2}\ell}{\kappa^{2}}\right)\lambda\left(1-\frac{\lambda}{2}\right)\left(1-\frac{\lambda}{4}\right)\ln\left(\frac{2L}{\ell}\right)\equiv\frac{c}{3}\ln\left(\frac{2L}{\ell}\right),\label{eq:exp_of_Eucl_act}\end{align}
 where $\frac{c}{3}$ is the coefficient of Weyl anomaly. Therefore,
the central charge reads\begin{equation}
c=\left(\frac{3\sigma}{4G_{3}\ell}\right)\left(2-\lambda\right),\label{eq:Weyl_coeff}\end{equation}
 which is exactly equal to the one obtained from Wald's formula (\ref{eq:BI-NMG_cent_chrg}).

\section{Conclusion}

We derived the equations of motion of the BI-NMG theory and discussed
the global AdS and asymptotically AdS solutions. We have shown that
assuming the null-energy condition, the theory admits simple $c$-functions
in the spirit introduced by Zamolodchikov. For asymptotically AdS
spaces, the value of the $c$-function at the boundary matches up
to a constant the central charge of the asymptotic symmetry algebra
and the coefficient of the Weyl anomaly. We have also shown that the
BI-NMG theory is free from the vacuum degeneracy problem of the NMG
theory. We still have not yet proven that BI-NMG theory has both a
unitary bulk and a unitary boundary theory, which was the problem
in NMG \cite{BHT} and the $O\left(R^{3}\right)$ extended NMG \cite{Sinha}.
After the completion of this manuscript, two related works \cite{Nam,Paulos}
have appeared. In \cite{Nam}, a detailed discussion on the AdS black
hole solutions of the $O\left(R^{3}\right)$ extension of \cite{Sinha}
and the BI-NMG theory as well as the computation of the central charge
are given. Some of our results overlap with \cite{Nam}. In \cite{Paulos},
an infinite order extension of NMG is discussed by requiring the existence
of a $c$-function. That $c$-function can be reproduced by expanding
(\ref{eq:inequlality_second}).

\section{\label{ackno} Acknowledgments}

T.Ç.\c{S}. is supported by a T{Ü}B\.{I}TAK PhD Scholarship. B.T.
is partially supported by the T{Ü}B\.{I}TAK Kariyer Grant No. 104T177.

\section*{Appendix A: $O\left(R^{4}\right)$ Expansion of the BI Action}

First, one should choose $\sigma=-1$. Using the expansion,\begin{align*}
\left[\det\left(1+A\right)\right]^{1/2} & =1+\frac{1}{2}\text{Tr}A-\frac{1}{4}\text{Tr}\left(A^{2}\right)+\frac{1}{8}\left(\text{Tr}A\right)^{2}+\frac{1}{6}\text{Tr}\left(A^{3}\right)-\frac{1}{8}\text{Tr}A\text{Tr}\left(A^{2}\right)+\frac{1}{48}\left(\text{Tr}A\right)^{3}\\
 & \phantom{=}-\frac{1}{8}\text{Tr}\left(A^{4}\right)+\frac{1}{32}\left[\text{Tr}\left(A^{2}\right)\right]^{2}+\frac{1}{12}\text{Tr}A\text{Tr}\left(A^{3}\right)-\frac{1}{32}\left(\text{Tr}A\right)^{2}\text{Tr}\left(A^{2}\right)+\frac{1}{384}\left(\text{Tr}A\right)^{4}+O\left(A^{5}\right).\end{align*}
 Then, $O\left(R^{4}\right)$ contributions to the Born-Infeld action
becomes\begin{align*}
O\left(R^{4}\right):\quad & \frac{1}{m^{8}}\left[-\frac{1}{8}\text{Tr}\left(g^{-1}Gg^{-1}Gg^{-1}Gg^{-1}G\right)+\frac{1}{32}\left[\text{Tr}\left(g^{-1}Gg^{-1}G\right)\right]^{2}+\frac{1}{12}\text{Tr}\left(g^{-1}Gg^{-1}Gg^{-1}G\right)\text{Tr}\left(g^{-1}G\right)\right.\\
 & \phantom{\frac{1}{m^{8}}}\left.-\frac{1}{32}\left[\text{Tr}\left(g^{-1}G\right)\right]^{2}\text{Tr}\left(g^{-1}Gg^{-1}G\right)+\frac{1}{384}\left[\text{Tr}\left(g^{-1}G\right)\right]^{4}\right]\\
 & =-\frac{1}{8m^{8}}\left[G^{\mu\nu}G_{\nu\alpha}G^{\alpha\beta}G_{\beta\mu}-\frac{1}{4}\left(G_{\mu\nu}^{2}\right)^{2}-\frac{2}{3}\left(G^{\mu\nu}G_{\nu\alpha}G_{\phantom{\beta}\mu}^{\alpha}\right)G_{\phantom{\beta}\beta}^{\beta}+\frac{1}{4}\left(G_{\phantom{\beta}\beta}^{\beta}\right)^{2}G_{\mu\nu}^{2}-\frac{1}{48}\left(G_{\phantom{\beta}\beta}^{\beta}\right)^{4}\right].\end{align*}
 Using (\ref{eq:Traces}) and \begin{align*}
G^{\mu\nu}G_{\nu\alpha}G^{\alpha\beta}G_{\beta\mu} & =R^{\mu\nu}R_{\nu\alpha}R^{\alpha\beta}R_{\beta\mu}-2R^{\mu\nu}R_{\nu\alpha}R_{\phantom{\alpha}\mu}^{\alpha}R+\frac{3}{2}R_{\mu\nu}^{2}R^{2}-\frac{5}{16}R^{4},\end{align*}
 one can obtain\begin{align*}
O\left(R^{4}\right):\quad & =-\frac{1}{8m^{8}}\left[R^{\mu\nu}R_{\nu\alpha}R^{\alpha\beta}R_{\beta\mu}-\frac{5}{3}R^{\mu\nu}R_{\nu\alpha}R_{\phantom{\alpha}\mu}^{\alpha}R+\frac{19}{16}R_{\mu\nu}^{2}R^{2}-\frac{1}{4}\left(R_{\mu\nu}^{2}\right)^{2}-\frac{169}{768}R^{4}\right].\end{align*}
 The coefficients of the above form solve the equations of consistency
for $O\left(R^{4}\right)$ extension in \cite{Sinha}.

\section*{Appendix B: Deriving the Equations of Motion}

In order to obtain equations of motion, let us consider the following
($\sigma=-1$) action\[
I=I_{\text{BI-NMG}}+I_{\text{matter}}.\]
 Then,\begin{align*}
\frac{\delta I_{\text{BI-NMG}}}{\delta g^{\mu\nu}} & =-\frac{4m^{2}}{\kappa^{2}}\int d^{3}x\,\sqrt{-\det g}\left[-\frac{1}{2}Fg_{\mu\nu}+F_{R}\left(\frac{\delta R}{\delta g^{\mu\nu}}-\frac{1}{m^{2}}\frac{\delta K}{\delta g^{\mu\nu}}-\frac{1}{6m^{4}}\frac{\delta S}{\delta g^{\mu\nu}}\right)\right],\end{align*}
 where $\frac{\partial F}{\partial R}$ is defined as $F_{R}$ and
given (\ref{eq:F_R_defn}). $\delta K$ reads as\begin{align*}
\delta K & =\delta R_{\mu\nu}^{2}-R\delta R\\
 & =\left(2R_{\nu}^{\phantom{\nu}\alpha}R_{\mu\alpha}-RR_{\mu\nu}\right)\delta g^{\mu\nu}-\left(2R_{\phantom{\alpha}\nu}^{\alpha}\nabla_{\mu}\nabla_{\alpha}-g_{\mu\nu}R^{\alpha\beta}\nabla_{\beta}\nabla_{\alpha}-R_{\mu\nu}\Box\right)\delta g^{\mu\nu}\\
 & \phantom{=}-R\left(g_{\mu\nu}\Box-\nabla_{\mu}\nabla_{\nu}\right)\delta g^{\mu\nu}.\end{align*}
 The $\delta S$ term is\begin{align*}
\delta S & =8\delta\left(R^{\mu\nu}R_{\mu\alpha}R_{\phantom{\alpha}\nu}^{\alpha}\right)-6\delta\left(RR_{\mu\nu}^{2}\right)+\delta\left(R^{3}\right)\\
 & =3\left[8R_{\phantom{\beta}\mu}^{\beta}R_{\beta\alpha}R_{\phantom{\alpha}\nu}^{\alpha}-4RR_{\nu}^{\phantom{\nu}\alpha}R_{\mu\alpha}+R_{\mu\nu}\left(R^{2}-2R_{\alpha\beta}^{2}\right)\right]\delta g^{\mu\nu}\\
 & \phantom{=}+12\left(g_{\mu\nu}R^{\alpha\rho}R_{\phantom{\beta}\rho}^{\beta}\nabla_{\beta}\nabla_{\alpha}+R_{\mu}^{\phantom{\mu}\rho}R_{\nu\rho}\Box-2R^{\alpha\rho}R_{\nu\rho}\nabla_{\mu}\nabla_{\alpha}\right)\delta g^{\mu\nu}\\
 & \phantom{=}-6R\left(g_{\mu\nu}R^{\alpha\beta}\nabla_{\beta}\nabla_{\alpha}+R_{\mu\nu}\Box-2R_{\phantom{\alpha}\nu}^{\alpha}\nabla_{\mu}\nabla_{\alpha}\right)\delta g^{\mu\nu}+3\left(R^{2}-2R_{\alpha\beta}^{2}\right)\left(g_{\mu\nu}\Box-\nabla_{\mu}\nabla_{\nu}\right)\delta g^{\mu\nu}.\end{align*}

Then, the equation of motion can be found as\begin{align*}
-\frac{\kappa^{2}}{8m^{2}}T_{\mu\nu} & =-\frac{1}{2}Fg_{\mu\nu}+\left(g_{\mu\nu}\Box-\nabla_{\mu}\nabla_{\nu}\right)F_{R}+F_{R}R_{\mu\nu}\\
 & \phantom{=}+\frac{1}{m^{2}}F_{R}\left[RR_{\mu\nu}-2R_{\lambda\nu\alpha\mu}R^{\lambda\alpha}-\Box\left(R_{\mu\nu}-\frac{1}{2}g_{\mu\nu}R\right)\right]\\
 & \phantom{=}+\frac{1}{m^{2}}\left[2\nabla_{\alpha}F_{R}\nabla_{\mu}R_{\phantom{\alpha}\nu}^{\alpha}-2\nabla_{\alpha}F_{R}\nabla^{\alpha}\left(R_{\mu\nu}-\frac{1}{2}g_{\mu\nu}R\right)-\nabla_{\nu}F_{R}\nabla_{\mu}R\right]\\
 & \phantom{=}+\frac{1}{m^{2}}\left[\left(2R_{\phantom{\alpha}\nu}^{\alpha}\nabla_{\alpha}\nabla_{\mu}-g_{\mu\nu}R^{\alpha\beta}\nabla_{\beta}\nabla_{\alpha}-R_{\mu\nu}\Box\right)+R\left(g_{\mu\nu}\Box-\nabla_{\mu}\nabla_{\nu}\right)\right]F_{R}\\
 & \phantom{=}-\frac{1}{m^{4}}F_{R}\left(2R_{\phantom{\rho}\mu}^{\rho}R_{\rho\alpha}R_{\phantom{\alpha}\nu}^{\alpha}+\left[g_{\mu\nu}\nabla_{\alpha}\nabla_{\beta}\left(R^{\beta\rho}R_{\phantom{\alpha}\rho}^{\alpha}\right)+\Box\left(R_{\nu}^{\phantom{\nu}\rho}R_{\mu\rho}\right)-2\nabla_{\alpha}\nabla_{\mu}\left(R_{\nu}^{\phantom{\nu}\rho}R_{\phantom{\alpha}\rho}^{\alpha}\right)\right]\right)\\
 & \phantom{=}-\frac{1}{2m^{4}}F_{R}\left(\left[2\nabla_{\alpha}\nabla_{\mu}\left(RR_{\phantom{\alpha}\nu}^{\alpha}\right)-g_{\mu\nu}\nabla_{\alpha}\nabla_{\beta}\left(RR^{\alpha\beta}\right)-\Box\left(RR_{\mu\nu}\right)\right]-2RR_{\nu}^{\phantom{\nu}\rho}R_{\mu\rho}\right)\\
 & \phantom{=}+\frac{1}{2m^{4}}F_{R}\left[\left(g_{\mu\nu}\Box-\nabla_{\nu}\nabla_{\mu}\right)+R_{\mu\nu}\right]\left(R_{\alpha\beta}^{2}-\frac{1}{2}R^{2}\right)\\
 & \phantom{=}-\frac{2}{m^{4}}\nabla_{\alpha}F_{R}\left[\nabla^{\alpha}\left(R_{\nu}^{\phantom{\nu}\rho}R_{\mu\rho}\right)+g_{\mu\nu}\nabla_{\beta}\left(R^{\beta\rho}R_{\phantom{\alpha}\rho}^{\alpha}\right)-\nabla_{\mu}\left(R_{\nu}^{\phantom{\nu}\rho}R_{\phantom{\alpha}\rho}^{\alpha}\right)\right]+\frac{2}{m^{4}}\nabla_{\alpha}\left(R_{\nu}^{\phantom{\nu}\rho}R_{\phantom{\alpha}\rho}^{\alpha}\right)\nabla_{\mu}F_{R}\\
 & \phantom{=}+\frac{1}{m^{4}}\nabla_{\alpha}F_{R}\left[\nabla^{\alpha}\left(RR_{\mu\nu}\right)+g_{\mu\nu}\nabla_{\beta}\left(RR^{\alpha\beta}\right)-\nabla_{\mu}\left(RR_{\phantom{\alpha}\nu}^{\alpha}\right)\right]-\frac{1}{m^{4}}\nabla_{\alpha}\left(RR_{\phantom{\alpha}\nu}^{\alpha}\right)\nabla_{\mu}F_{R}\\
 & \phantom{=}-\frac{1}{2m^{4}}\left(\nabla_{\mu}F_{R}\nabla_{\nu}+\nabla_{\nu}F_{R}\nabla_{\mu}-2g_{\mu\nu}\nabla_{\rho}F_{R}\nabla^{\rho}\right)\left(R_{\alpha\beta}^{2}-\frac{1}{2}R^{2}\right)\\
 & \phantom{=}-\frac{1}{m^{4}}\left(g_{\mu\nu}R^{\beta\rho}R_{\phantom{\alpha}\rho}^{\alpha}\nabla_{\alpha}\nabla_{\beta}+R_{\nu}^{\phantom{\nu}\rho}R_{\mu\rho}\Box-2R_{\nu}^{\phantom{\nu}\rho}R_{\phantom{\alpha}\rho}^{\alpha}\nabla_{\alpha}\nabla_{\mu}\right)F_{R}\\
 & \phantom{=}-\frac{1}{2m^{4}}\left(2RR_{\phantom{\alpha}\nu}^{\alpha}\nabla_{\alpha}\nabla_{\mu}-g_{\mu\nu}RR^{\alpha\beta}\nabla_{\alpha}\nabla_{\beta}-RR_{\mu\nu}\Box\right)F_{R}\\
 & \phantom{=}+\frac{1}{2m^{4}}\left(R_{\alpha\beta}^{2}-\frac{1}{2}R^{2}\right)\left(g_{\mu\nu}\Box-\nabla_{\nu}\nabla_{\mu}\right)F_{R},\end{align*}
 where \begin{align*}
\nabla_{\alpha}\nabla_{\mu}R_{\phantom{\alpha}\nu}^{\alpha} & =\frac{1}{2}\nabla_{\mu}\nabla_{\nu}R+R_{\lambda\mu}R_{\phantom{\lambda}\nu}^{\lambda}-R_{\lambda\nu\alpha\mu}R^{\lambda\alpha}\end{align*}
 is used. This form is more suitable for the linearization and the
conserved charge analysis which we shall return in a separate work.

The trace of (\ref{eq:Eqn_of_motn}) is also of some use\begin{align*}
-\frac{\kappa^{2}}{8m^{2}}T & =-\frac{3}{2}F+2\Box F_{R}+F_{R}R\\
 & \phantom{=}-\frac{1}{m^{2}}\left[\nabla_{\beta}\nabla_{\alpha}\left(F_{R}R^{\alpha\beta}\right)+2F_{R}R_{\mu\nu}^{2}-\Box\left(F_{R}R\right)-F_{R}R^{2}\right]\\
 & \phantom{=}-\frac{1}{2m^{4}}\left[4F_{R}R_{\phantom{\rho}\mu}^{\rho}R_{\rho\alpha}R^{\alpha\mu}+2\nabla_{\alpha}\nabla_{\beta}\left(F_{R}R^{\beta\rho}R_{\phantom{\alpha}\rho}^{\alpha}\right)-\nabla_{\alpha}\nabla_{\beta}\left(F_{R}RR^{\alpha\beta}\right)-3F_{R}RR_{\mu\nu}^{2}+\frac{1}{2}F_{R}R^{3}\right].\end{align*}

\end{document}